\documentclass[a4paper]{jpconf}
\usepackage{graphicx}
\usepackage{amsmath,amssymb,amsfonts}
\usepackage{subfigure}
\usepackage{listings}
\usepackage{cite}

\newcommand{\secdec}{{\textsc{SecDec}}}
\def\be{\begin{equation}}
\def\ee{\end{equation}}
\newcommand{\bea}{\begin{eqnarray}}
\newcommand{\eea}{\end{eqnarray}\noindent}
\newcommand{\nn}{\nonumber}
\newcommand{\bcen}{\begin{center}}
\newcommand{\ecen}{\end{center}}

\def\url#1{\texttt{#1}}
\def\eps{\epsilon}

\begin{document}
\title{Numerical multi-loop calculations with \textsc{SecDec}}

\author{Sophia Borowka, Gudrun Heinrich\footnote{Speaker; presented at the conference ACAT 2013,
Beijing, China, May 2013.}}
\address{Max Planck Institute for Physics, F\"ohringer Ring 6, 80805 Munich, Germany}


\begin{abstract}
The new version 2.1 of the program \secdec{} is described, which can be used for the 
factorisation of poles and subsequent 
numerical evaluation of multi-loop integrals, in particular massive two-loop integrals.
The program is not restricted to scalar master integrals;
more general parametric integrals can also be treated in an automated way.
\end{abstract}

\section{Introduction}

In the absence of smoking gun signals of New Physics at the LHC so far, 
precision measurements play a crucial role in the search for 
less direct signs of New Physics and in exploring the Higgs sector. 
Certainly this is only possible if  precise theory predictions
are available. 
In most cases, this means that next-to-leading order (NLO) 
calculations are necessary, ideally matched to a parton shower. 
However, there is a number of examples where NNLO precision is required.
While in the case of jet production the 
main challenge  at NNLO consists in the treatment of the doubly 
unresolved real radiation part, 
processes involving massive particles --  
e.g. top quarks or $W/Z$ bosons -- contain another bottleneck, 
given by the two-loop integrals entering the virtual corrections.
The recent result for $t\bar{t}$ production at NNLO\,\cite{Czakon:2013vfa}
is based on a semi-numerical representation of the integrals\,\cite{Czakon:2008zk}, 
while fully analytical representations for most of the master integrals needed 
for this process have been worked out 
in \cite{vonManteuffel:2013uoa,vonManteuffel:2012je,Bonciani:2010mn,Bonciani:2009nb,Bonciani:2008az}.
Planar two-loop master integrals entering the production of two massive vector bosons at NNLO 
have been calculated in \cite{Gehrmann:2013cxs}, and in the high energy limit 
in \cite{Chachamis:2008yb} for the case of $W$ pair production. 

However, non-planar massive two-loop integrals which cannot be expressed entirely in terms of 
generalized harmonic polylogarithms still pose a problem for a fully analytical evaluation. 
Numerical methods on the other hand do not have these difficulties. For the latter the problem rather 
lies in the isolation of the singularities, as well as in numerical efficiency and accuracy.

The program \secdec{}, presented in \cite{Carter:2010hi,Borowka:2012yc}, performs the task of 
isolating singularities which, if regulated by dimensional regularisation, 
appear as poles in $1/\eps$, in an automated way, 
based on the algorithm of 
sector decomposition \cite{Binoth:2000ps,Roth:1996pd,Hepp:1966eg}.
Other implementations of sector decomposition into public programs are also available
\cite{Smirnov:2009pb,Bogner:2007cr,Gluza:2010rn}. 
However, the latter are more or less restricted to the Euclidean region, 
while \secdec{} can deal with physical kinematics.

In this talk the  new features of the 
program \secdec\,2.1 will be presented, along with some results.

\section{Structure, usage and new features of \secdec{} version 2.1}
\label{sec:install}

\subsection{Structure}

\begin{figure}[htb]
\includegraphics[width=14cm]{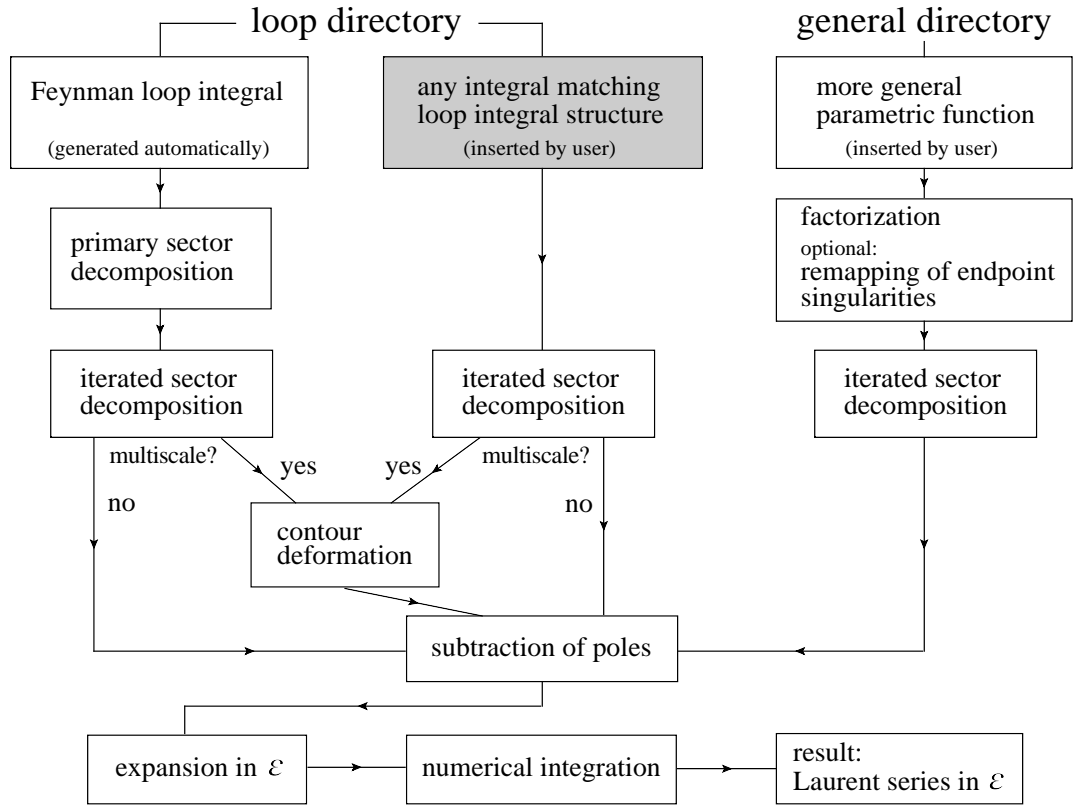}
\caption{Flowchart showing the structure of the program \secdec{}.}
\label{fig:flowchart}
\end{figure}

As can be seen from Fig.~\ref{fig:flowchart}, the program contains two 
sections, one for loop integrals, one for more general parametric functions
(corresponding to the directories {\it loop} and {\it general}).
The {\it loop} part has been extended in version 2.1 to be able to treat 
parametric integrals which are not in the canonical form as  
obtained directly 
from standard Feynman parametrisation. They can have a different 
format coming for example from variable transformations and/or  analytical 
integrations over some Feynman parameters. 
Hence, as these functions differ from the standard representation which the program would derive from 
the propagators in an automated way, they have to be defined in an input file by the user.
The detailed syntax to define these functions is given in subsection \ref{sec:userdefined}.
Contour deformation is available for these functions, as they are assumed to originate 
from a Feynman integral structure where poles on the real axis are protected 
by the infinitesimal $i\delta$ prescription.

The part in the directory {\it general} can deal with even more general parametric
functions, where the integrand can consist of a product 
of arbitrary length of polynomial functions to some power. 
However, these functions should have only endpoint singularities
(i.e. dimensionally regulated singularities at the integration boundaries). 
Contour deformation is not available in this case because the 
correct sign of the imaginary part steering the deformation into the complex plane 
cannot be inferred if the assumption of an underlying 
loop integral structure is dropped.

The procedure of iterated sector decomposition and subsequent treatment 
is the same for all the different types of input functions, 
and is described in \cite{Carter:2010hi,Borowka:2012yc}.

\subsection{Installation}
The current version of the program can be downloaded from \\
{\tt http://projects.hepforge.org/secdec}.

Unpacking the tar archive via 
{\it  tar xzvf SecDec-2.1.2.tar.gz} 
will create a directory called {\tt SecDec-2.1.2}. 
Changing to the {\tt SecDec-2.1.2} directory
and running {\it ./install} will compile the {\sc Cuba} library\,\cite{Hahn:2004fe} 
needed for the numerical integration. 

Prerequisites are Mathematica, version 6 or above, Perl (installed by default on 
most Unix/Linux systems) and  a C++ compiler.
If the   Fortran option is used, a Fortran compiler 
is obviously  also required.

\subsection{Usage}

There are two files to be edited: a text file {\tt param*.input} where the parameters are
set, and a Mathematica file {\tt template*.m} where the funtion(s) to be integrated
are defined. In more detail:
\begin{enumerate}
\item 
In the {\tt loop} directory, edit the files 
{\tt paramloop.input} and {\tt templateloop.m} if you want to compute a Feynman loop
integral in a fully automated way.
If you would like to define a set of own functions rather than a standard loop integral, 
take the files {\tt paramuserdefined.input} and {\tt templateuserdefined.m} 
as a starting point.
In the {\tt general} directory, the analogous files are {\tt param.input} and 
{\tt template.m}.

Let us call the files edited by the user {\tt myparamfile.input} and {\tt mytemplatefile.m}.

\item Execute the command {\it ./launch -p myparamfile.input -t mytemplatefile.m} 
in the shell. If you add the option {\it -u } in the {\tt loop} directory, user defined functions are computed. \\
If your files {\it myparamfile.input, mytemplatefile.m} are in a different directory, say, 
{\it myworkingdir}, 
 use the option {\bf -d myworkingdir}, i.e. the full command then looks like 
 {\it ./launch -d myworkingdir -p myparamfile.input -t mytemplatefile.m}, 
 executed from the directory {\tt SecDec/loop} or
 {\tt SecDec/general}.
 \item Collect the results. Depending on whether you have used a single machine or 
submitted the jobs to a cluster, the following actions will be performed:
 \begin{itemize}
\item If the calculations are done sequentially on a single machine, 
    the results will be collected automatically (via the corresponding {\tt results*.pl} called by {\tt launch}).
    The output file will be displayed with your specified text editor.
\item If the jobs have been submitted to a cluster,    
	when all jobs have finished, execute the command 
	{\it ./results.pl [-d myworkingdir -p myparamfile]} in the general, and 
	{\it ./resultsloop.pl [-d myworkingdir -p myparamfile]} or
	{\it ./resultsuserdefined.pl [-d myworkingdir -p myparamfile]} in the loop directory, respectively.
	This will create the files containing the final results in the {\tt graph} subdirectory
	specified in the input file.
\end{itemize}
\item After the calculation and the collection of the results is completed, 
you can use the shell command {\it ./launchclean[graph]}
to remove obsolete files.
\end{enumerate}

More details about the usage can be found in Ref.~\cite{Borowka:2013cma}, 
and also in the documentation coming with the program.

\subsection{New Features}
Version 2.1 of  \secdec{} contains the following new features, 
also partially illustrated by examples in Section \ref{sec:results}. 

\subsubsection{Evaluation of user-defined functions in the loop part:}
\label{sec:userdefined}


If the user would like to calculate a ``standard" loop integral, it is sufficient 
to specify the propagators, and the program will construct the integrand in terms of 
Feynman parameters automatically.
However, analytical  manipulations on the integrand before starting the 
iterated sctor decomposition  can be helpful when dealing with 
complicated integrals. 
For example, integrating out one Feynman parameter analy\-ti\-cally reduces the number of integration variables for the subsequent
Monte Carlo integration and therefore can improve numerical efficiency. 
This implies that the constraint $\delta(1-\sum_i x_i)$ has been used already to achieve a 
convenient parametrisation, and therefore no primary sector decomposition 
to eliminate the $\delta$-constraint is needed anymore.
In such a case, the user can skip the primary sector decomposition step and 
insert the integrand functions  directly into the Mathematica input file.

The detailed syntax is as follows.
%
The user-defined functions should be polynomial in the Feynman parameters
and can also involve kinematic invariants, i.e. 
should be structurally similar to the
functions of type  $\mathcal{U}$ and  $\mathcal{F}$ (see e.g. eq.~(\ref{GNP})).
They can be raised to some  power 
which does not have to be an integer.  
These functions in addition can be multiplied by an arbitrary 
``numerator" function. The only condition the latter must fulfill is that it 
should not contain any singularities or thresholds.
In case the functions $\mathcal{U}$ and $\mathcal{F}$ contain thresholds, 
a deformation of the integration contour into the complex plane will be performed. 
The setup is such that
the integration contour will be formed based on the function $\mathcal{F}$.
 The list of user-defined functions must be inserted into  {\tt mytemplatefile.m} using the following syntax\\
\texttt{functionlist}=\{{\it function\_1},{\it function\_2},...,{\it function\_i},...\};\\
  with 
{\it function\_i}=\{{\it \# of function},\{{\it list of exponents}\},\{\{{\it function $\mathcal{U}$}, {\it exponent of $\mathcal{U}$}, {\it decomposition flag}\},\{{\it function $\mathcal{F}$},{\it exponent of $\mathcal{F}$},{\it decomposition flag}\}\}, {\it numerator}\}.\\
The {\it \# of function} is an integer labelling the function, 
where the default is sequential numbering.

Each entry in the comma separated {\it list of exponents} corresponds to an exponent of a 
Feynman parameter occurring as a monomial in the Feynman integral.
The {\it decomposition flag} should be  {\tt A} if no iterated sector decomposition is desired, 
and  {\tt B} if the function needs further decomposition. 
The {\it numerator} may contain several functions, with different exponents,  
as long as the functions are non-singular.
An example including detailed comments can be found in {\tt /loop/templateuserdefined.m} and 
{\tt /loop/paramuserdefined.input}.

\subsubsection{Tensor integrals:}


For the computation of tensor integrals, where the tensor is contracted with external momenta and/or loop momenta, 
the construction of the Feynman integral via topological cuts needs to be switched off, 
which corresponds to {\tt cutconstruct=0}. \\
To define the numerator in  {\tt mytemplatefile.m}, 
each scalar product of loop momenta contracted with either external momenta or 
loop momenta should be given as an entry of a list:\\
{\tt numerator=\{{\it prefactor},{\it comma separated list of scalar products}\}}.
For example, a numerator of the form  $2\,(k_1\cdot p_1)(k_1\cdot k_2)$, 
where $p_1$ is an external momentum and the $k_i$ are loop momenta, 
should be given as 
{\tt numerator=\{}2, $k1*p1$, $k1*k2$ {\tt \}}.

\subsubsection{Error estimates:} 

\medskip

When dealing with complicated integrands it can happen that the error given by 
the Monte Carlo integration program -- which is based on 
the number of sampling points only -- underestimates the true error.
The numerical integrators contained in the {\scshape Cuba} library~\cite{Hahn:2004fe,Agrawal:2011tm} 
give an estimate how trustworthy
the stated error is.
The new \secdec{} version 2.1 collects the maximal error probability for each computed 
order in the Laurent series in $\epsilon$ and
 writes it to the result files {\tt *.res}. 
In the generic case, the information on the reliability of 
the stated error is given as a probability with values between 0 and 1. 
If the integrator returns a value larger than one, 
the integration has not come to  successful completion, 
and a warning is written to the result files. 
This feature should enable the user to assess the reliability of the numerical result, 
respectively the given error estimate.

\section{Examples and Results}
\label{sec:results}

\subsection{Non-planar massive two-loop diagrams entering NNLO $t\bar{t}$ production}

The sector decomposition algorithm 
is described in detail in~\cite{Binoth:2000ps,Heinrich:2008si}, 
while the general structure of the program \secdec{} is described in~\cite{Carter:2010hi,Borowka:2012yc}.
Therefore we concentrate mainly on the new features and results here.

\vspace*{3mm}

\begin{figure}[h]
\subfigure[ggtt1]{\raisebox{8pt}{\includegraphics[width=0.4\textwidth]{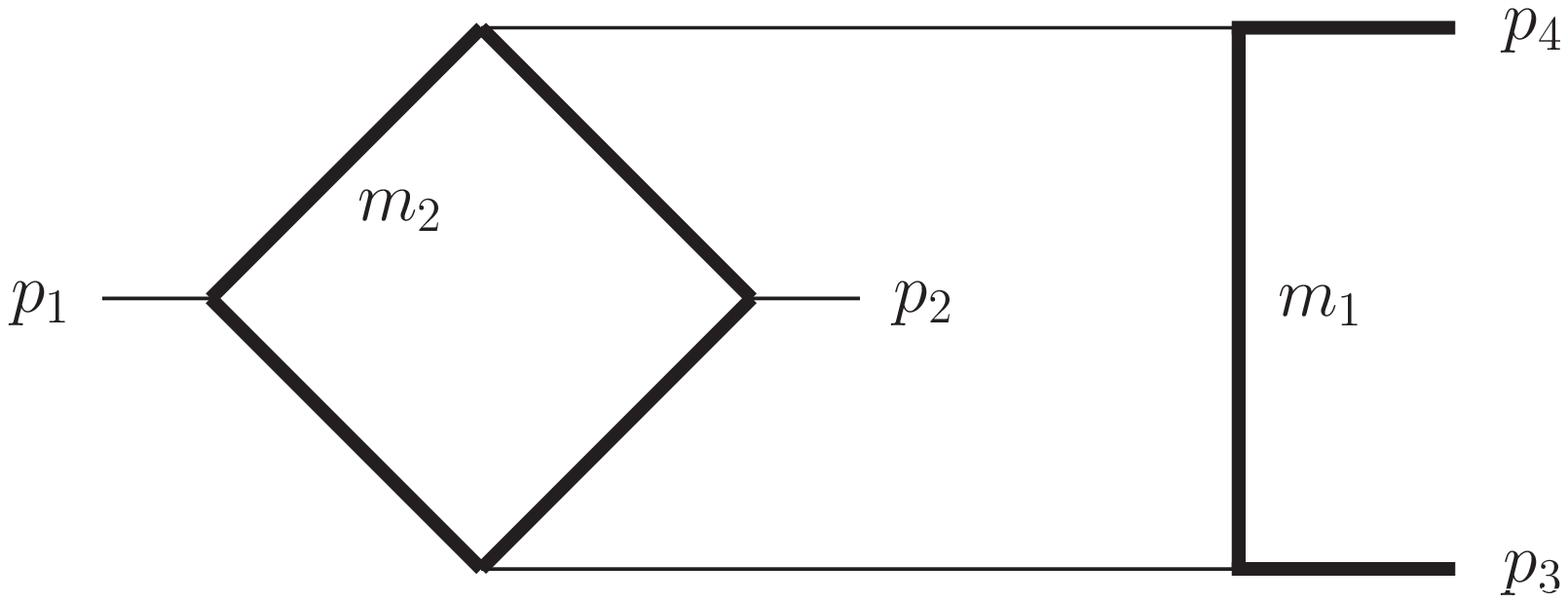}} }
\hspace{2cm}
\subfigure[ggtt2]{\includegraphics[width=0.35\textwidth]{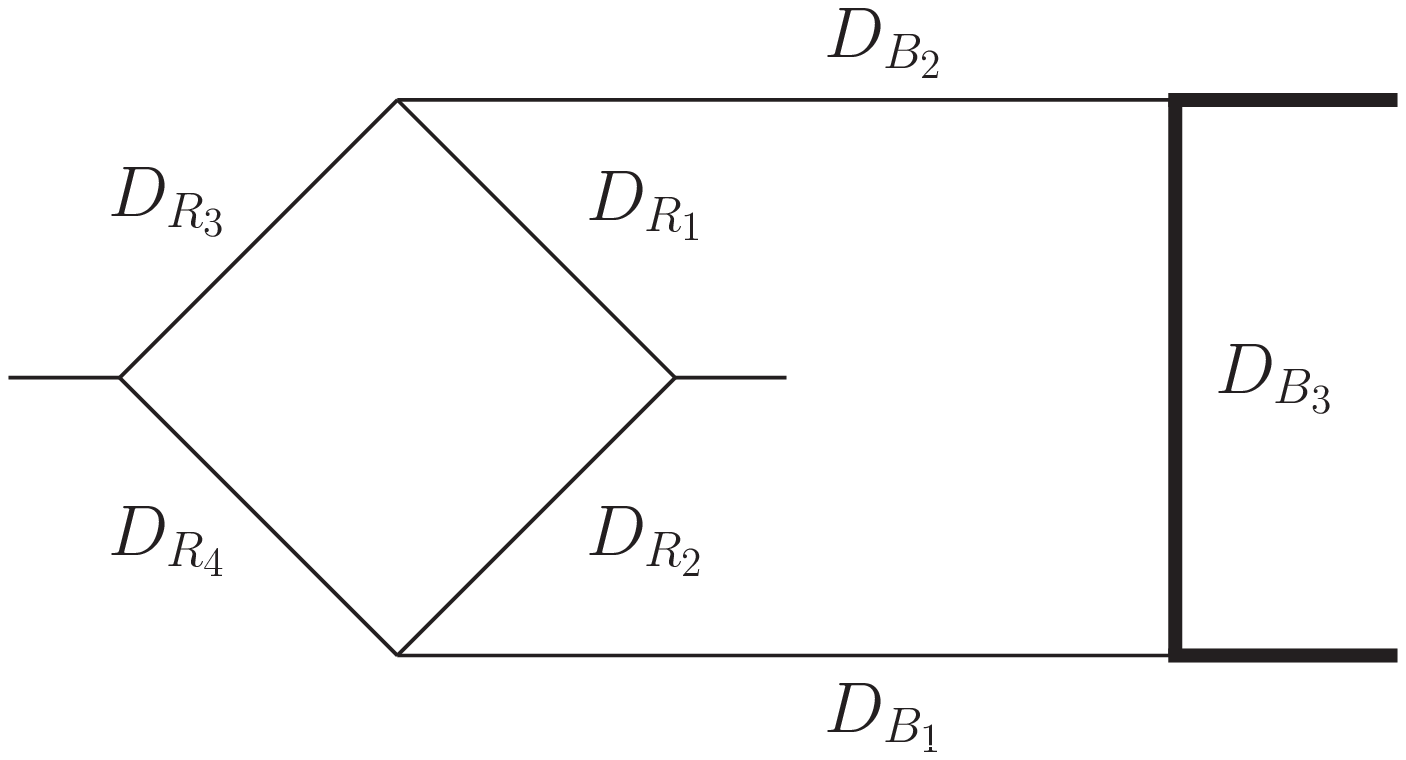} }
	\caption{Massive non-planar two-loop box diagrams entering the heavy (a) and 
	light (b) fermionic correction to the $gg\to t\bar{t}$ channel; the bold lines denote massive particles.} 
	\label{fig:ggtt2} 
\end{figure}

The most complicated master topologies occurring in the two-loop corrections 
to $t\bar{t}$ production in the $gg$ channel are the non-planar seven-propagator integrals shown in 
Fig.\ref{fig:ggtt2}.

Analytic results for the integral 
containing a sub-diagram with a massive loop (called {\it ggtt1} here), are not available, 
while analytic results for 
the diagram corresponding to massless fermionic corrections 
in a sub-loop (called {\it ggtt2} here) have become available only 
very recently\,\cite{vonManteuffel:2013uoa}.
However, the numerical evaluation of {\it ggtt1} with \secdec{} is 
much easier than the one of {\it ggtt2}, 
due to its less complicated infrared singularity structure. 
While the leading poles of {\it ggtt2} are of order $1/\eps^{4}$, 
and intermediate expressions during sector decomposition 
contain (spurious) poles where the degree of divergence is higher than logarithmic,
the integral {\it ggtt1} is finite and free from spurious poles.
Therefore we can evaluate {\it ggtt1} with \secdec\,2.1 
using the fully automated setup.
In contrast,  for {\it ggtt2}
it turned out to be advantageous to make some analytical manipulations beforehand.
In particular, it was useful to perform one parameter integration analytically 
before feeding the integral into the decomposition and numerical integration algorithm. 
Further, we introduced special transformations to reduce the occurrence of spurious 
singularities, which are described in detail in \cite{Borowka:2013cma}.
These manipulations lead to functions which were not in the ``standard form" 
of Feynman parameterised loop integrals   anymore. 
Therefore this entailed the development of a setup for ``non-standard" 
parametric functions, which has been made available for the user as 
described in section \ref{sec:userdefined}, 
as it can be beneficial in other contexts as well.

The expression for the scalar integral {\it ggtt2}  in momentum space is given by
\begin{align}
\label{eq:gnp}
 \mathcal{G}_{ggtt2}= \left( \frac{1}{\mathrm{i} \pi^{\frac{\mathrm{D}}{2}}}\right)^2 
 \int \frac{\mathrm{d^D}k_1\,\, \mathrm{d^D}k_2}{D_{R_1}D_{R_2}D_{R_3}D_{R_4}D_{B_1}D_{B_2}D_{B_3}}  
\end{align}
where $\mathrm{D}=4-2 \epsilon$.
Integrating out the loop momentum $k_2$ (corresponding to massless sub-loop) first, 
we are left with an expression containing 
$k_1$ and external momenta only:
\begin{align}
 \mathcal{I}_{R} = \frac{1}{\mathrm{i} \pi^{\mathrm{D}/2}} \int \frac{ \mathrm{d^D}k_2 }
 { D_{R_1}D_{R_2}D_{R_3}D_{R_4}}
 = \Gamma(2+\epsilon) \int \prod_{i=1}^4\mathrm{d}x_i \delta(1-\sum_{j=1}^4 x_j) 
 \mathcal{F}(\vec{x},k_1)^{-2-\epsilon} \text{ ,}\label{loop1}
\end{align}
with
\begin{align}
\!\! -\mathcal{F}(\vec{x},k_1)= (k_1-p_3)^2 x_1x_2 + (k_1+p_1+p_4)^2x_1x_3 + 
(k_1+p_2+p_4)^2 x_2x_4 +  (k_1+p_4)^2 x_3x_4 \text{ .}\nn
\end{align}
We eliminate the $\delta$-function in eq.\,(\ref{loop1}) with the substitution
\begin{align}
\label{eq:parametrisation}
x_1 = t_2 (1- t_3) \text{ ,}\hspace{19pt} x_2 = t_1 t_3 \text{ ,}\hspace{19pt} x_3 = (1- t_1) t_3 \text{ ,}
\hspace{19pt}
\end{align}
which allows us to integrate out the parameter $t_3$ analytically.
Then we combine the expression for  $\mathcal{I}_{R}$ 
with the remaining $k_1$-dependent propagators 
to obtain, after integrating out $k_1$, 
\begin{align}
\nonumber \mathcal{G}_{NP}=&\frac{2}{\epsilon}
\frac{\Gamma(3+2\epsilon)\Gamma^2(1-\epsilon)}{\Gamma(1-2\epsilon)} 
\int_0^1 \mathrm{d}t_1\int_0^1\mathrm{d}t_2 \,\, \times\label{GNP} \\
 &\prod_{i=1}^4\int_0^1 \mathrm{d}z_i \,z_4^{1+\epsilon} \,\delta(1-\sum_{j=1}^4 z_j) 
 \,\,\mathcal{F_{NP}}(\vec{z},t_1,t_2)^{-3-2\epsilon} \,\,
 \mathcal{U_{NP}}(\vec{z})^{1+3\epsilon} \text{ ,}
\end{align}
\begin{eqnarray}\label{eq:Fnp}
&&\mathcal{U_{NP}}(\vec{z})= \sum_{j=1}^4 z_j \;, \;
\mathcal{F_{NP}}(\vec{z},t_i)= 
-s_{12} z_2 z_3 - T z_1 z_4 - S_1 z_2 z_4 -S_2 z_3 z_4 + m^2 z_1 (z_1+z_4 Q)  \text{ ,}\nonumber\\
&&T = s_{13} \bar{t}_1 t_2 +s_{23} t_1 \bar{t}_2 \text{ ,}\;
S_1= s_{12} t_1 t_2 \text{ ,}\; S_2=s_{12} \bar{t}_1 \bar{t}_2 \;,\;
 Q =t_1\bar{t}_2+\bar{t}_1t_2\text{ ,}\; s_{ij} = (p_i+p_j)^2\text{ .}
\end{eqnarray}
We use the shorthand notation $\bar{t}_i=1-t_i$.
Now we eliminate the $\delta$-function in eq.~(\ref{GNP}) by performing a 
primary sector decomposition~\cite{Binoth:2000ps} in $z_1,\dots,z_4$.

Note that, due to the substitutions made in (\ref{eq:parametrisation}),  
the integral $\mathcal{G}_{NP}$ can have singularities both at zero and one in $t_1$ and $t_2$. 
With the sector decomposition algorithm, only singularities at zero are factorised automatically. 
Consequently, we remap the singularities located at the upper integration limit 
to the origin of parameter space by splitting the integration region at $\tfrac{1}{2}$
and transforming the integration variables to remap the integration domain to the unit cube\,\cite{Heinrich:2008si}.
This procedure results in 12 integrals,   
some of which already being finite, such that no subsequent sector decomposition is required.
Some of the integrals however lead to singularities of the type $\int_0^1dx\,x^{-2-\eps}$ after sector decomposition, 
which we call {\it linear } divergences. These singularities are spurious, so it would be  
better to avoid this type of singularity from scratch.
In the following section  we sketch a method which has proven useful in this respect\,\cite{Borowka:2013cma}. 

\subsubsection{Removal of double linear divergences via backwards transformation:}
\label{subsec:backsecdec}
To explain the type of transformation advocated here,  
we use as an example the following function, which is part of eq.~(\ref{GNP}) after 
primary sector decomposition
\begin{align}
\mathcal{F}(\vec{z})&= -s_{12} z_3 - t\, z_1 z_4 - s_1\, z_4 -s_2\, z_3 z_4 + m^2 z_1 (z_1 +z_4
(z_2\bar{z}_5+z_5 \bar{z}_2))\nonumber\\
t&=s_{13} \bar{z}_5 z_2 +s_{23} z_5 \bar{z}_2 \text{ ,}\;
s_1= s_{12} z_5 z_2 \text{ ,}\; s_2=s_{12} \bar{z}_5 \bar{z}_2 \;.
\label{eq:Fsec2}
\end{align}
Concerning the Feynman parameters, we can identify the following structure in 
eq.~(\ref{eq:Fsec2}):
\begin{align}
\mathcal{F}(\vec{z})& = 
 z_4\,\Big( P(\vec{z}_{14}) + z_1 Q(\vec{z}_{14}) \Big) + R(\vec{z}_{14})  \text{ ,}
\label{eq:secback}
\end{align}
where $P,Q$ and $R$ are 
polynomials of arbitrary degree of  Feynman parameters and kinematic invariants, with 
$\vec{z}_{14}=\{z_2,z_3,z_5\}$, i.e. $P,Q$ and $R$ do not depend on $z_1$ and $z_4$. 
In eq.~(\ref{eq:secback}), all terms multiplied by the Feynman parameter $z_1$ are 
also multiplied by the Feynman parameter $z_4$. Hence, the sector decomposition method can be applied ``backwards".
To explain this in more detail, consider the following function:
\begin{align}
J=&\int_0^1  {\rm d}z_1\ldots {\rm d}z_5 \,z_4^{-1-\eps}
\left[ z_4 P + z_1 Q + R\right]^{-3-2\eps} [\underbrace{\Theta(z_1-z_4)}_{(1)}+\underbrace{\Theta(z_4-z_1)}_{(2)}]\label{eq:wayback}
\end{align}
Now we substitute $z_4=z_1\,t_4$ in sector (1) and  $z_1=z_4\,t_1$ in sector (2), to obtain, 
after renaming again $t_i$ into $z_i$
\begin{align}
J &= \int_0^1  {\rm d}z_1\ldots {\rm d}z_5 \,z_4^{-1-\eps}
\left[ z_4 P + z_1 Q + R\right]^{-3-2\eps} \label{eq:backwards0}\\
&=\int_0^1  {\rm d}z_1\ldots {\rm d}z_5 \, z_4^{-1-\eps}z_1^{-\eps}
\left[ z_1 (z_4 P + Q ) + R\right]^{-3-2\eps} \label{eq:backwards1}\\
&\quad+\int_0^1  {\rm d}z_1\ldots {\rm d}z_5 
\left[ z_4 ( P + z_1 Q ) + R\right]^{-3-2\eps} \;.\label{eq:backwards2}
\end{align}
We observe that the  term in (\ref{eq:backwards2}) is the same as in eq.~(\ref{eq:secback}). 
Therefore we can replace it by the expressions (\ref{eq:backwards0}) minus (\ref{eq:backwards1}). 
Doing so, the effect is twofold: The degree of the polynomial in $(z_1z_4)$  is reduced  in eq.~(\ref{eq:backwards0}),  
and  in eq.~(\ref{eq:backwards1})  
$z_4$ is traded for $z_1$ to multiply $P$ and $Q$, which can be beneficial in view of further decompositions. 

After all transformations of this type we arrive at a total of 15 functions partly needing an iterated sector decomposition. 
Making use of the new feature of {\it user-defined functions} 
described above, we were able to compute the {\it ggtt2} diagram much more efficiently than with the 
standard setup.

\subsubsection{The {\it ggtt2} diagram:}

\begin{figure}[htb]
\subfigure[finite part, including threshold]{
\includegraphics[width=7cm]{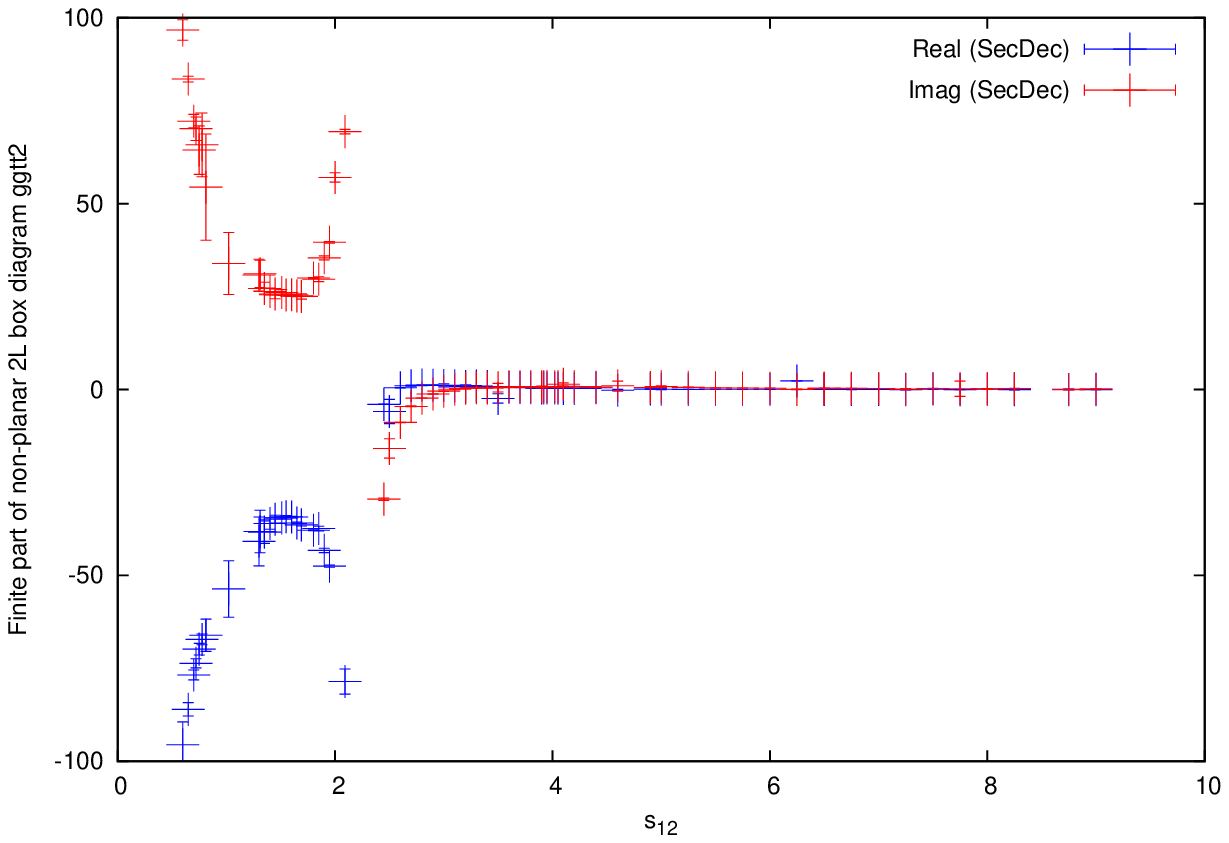} }
\hfill
\subfigure[finite part, beyond threshold]{
\includegraphics[width=7cm]{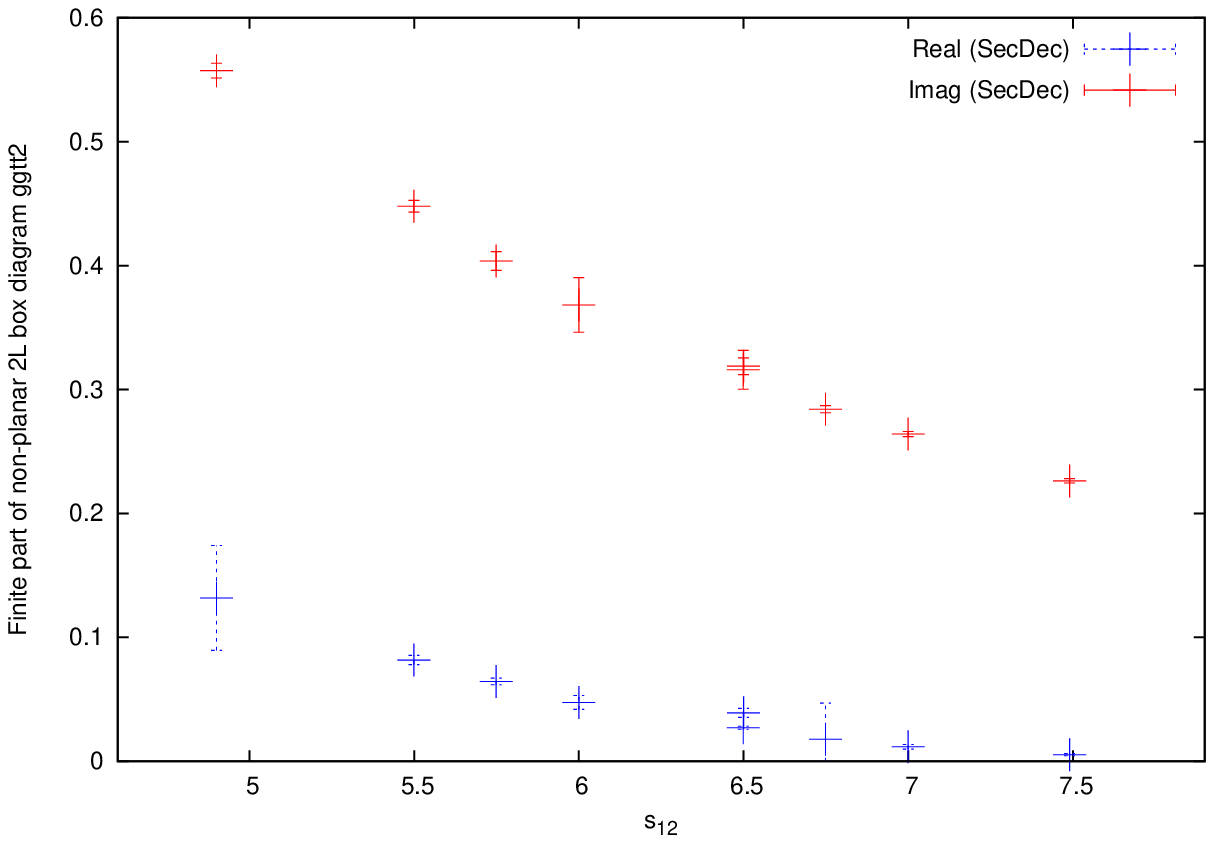} }
\caption{Results for the finite part of the scalar integral  {\it ggtt2}, 
(a) for a larger kinematic range, (b) zoom into a region further away from threshold.
 The vertical bars denote the numerical integration errors.}
\label{fig:ggtt2finite}
\end{figure}

Numerical results for the {\it ggtt2} diagram are shown in 
Fig.~\ref{fig:ggtt2finite}, 
where we used 
the numerical values $p_3^2=p_4^2=m^2=1, s_{23}=-1.25, s_{13}=2m^2-s_{12}-s_{23}$, 
and we extract an overall factor of $-16\,\,\Gamma(1+\eps)^2$. 
We only show results for the  finite part here, 
as it is the most complicated one and therefore more interesting than the pole coefficients.

\subsubsection{The {\it ggtt1} diagram:}

\begin{figure}[htb]
\subfigure[scalar integral]{\includegraphics[width=7.cm]{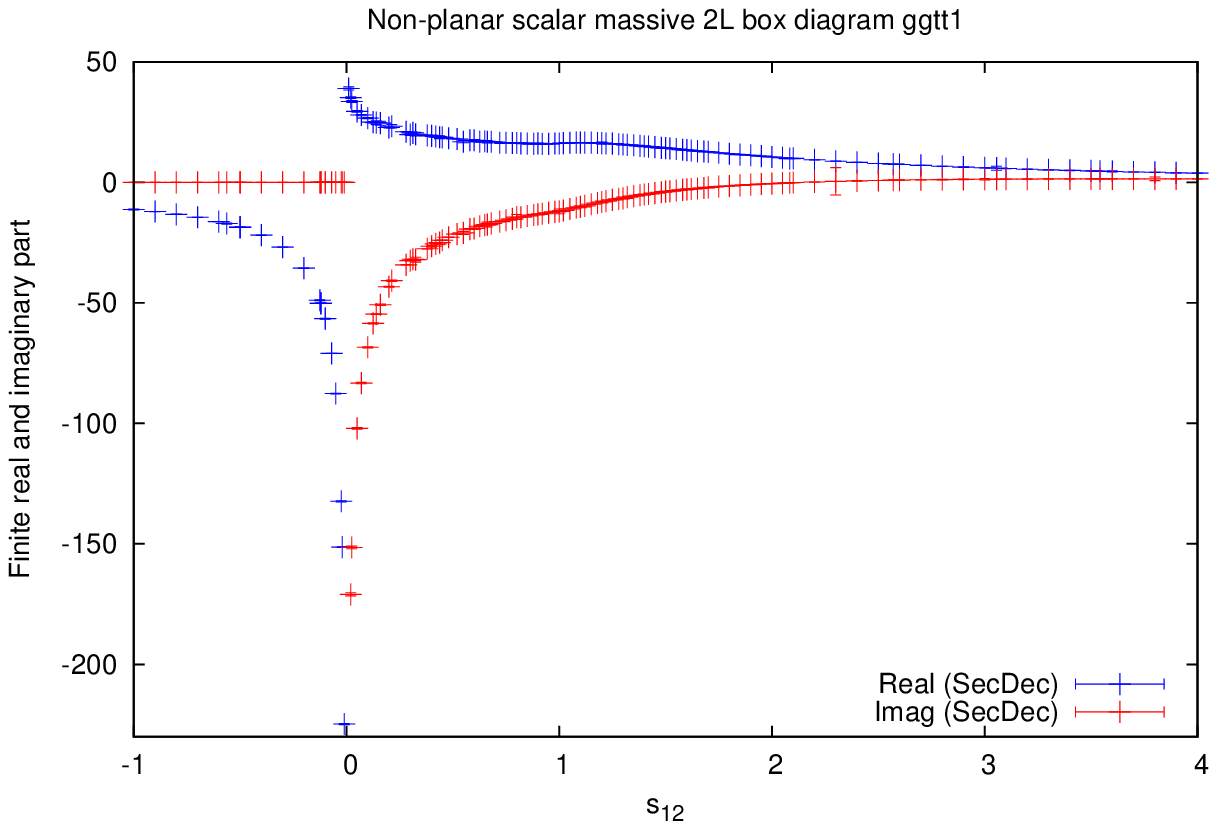} }\hfill
\subfigure[rank 2 tensor integral]{\includegraphics[width=7.cm]{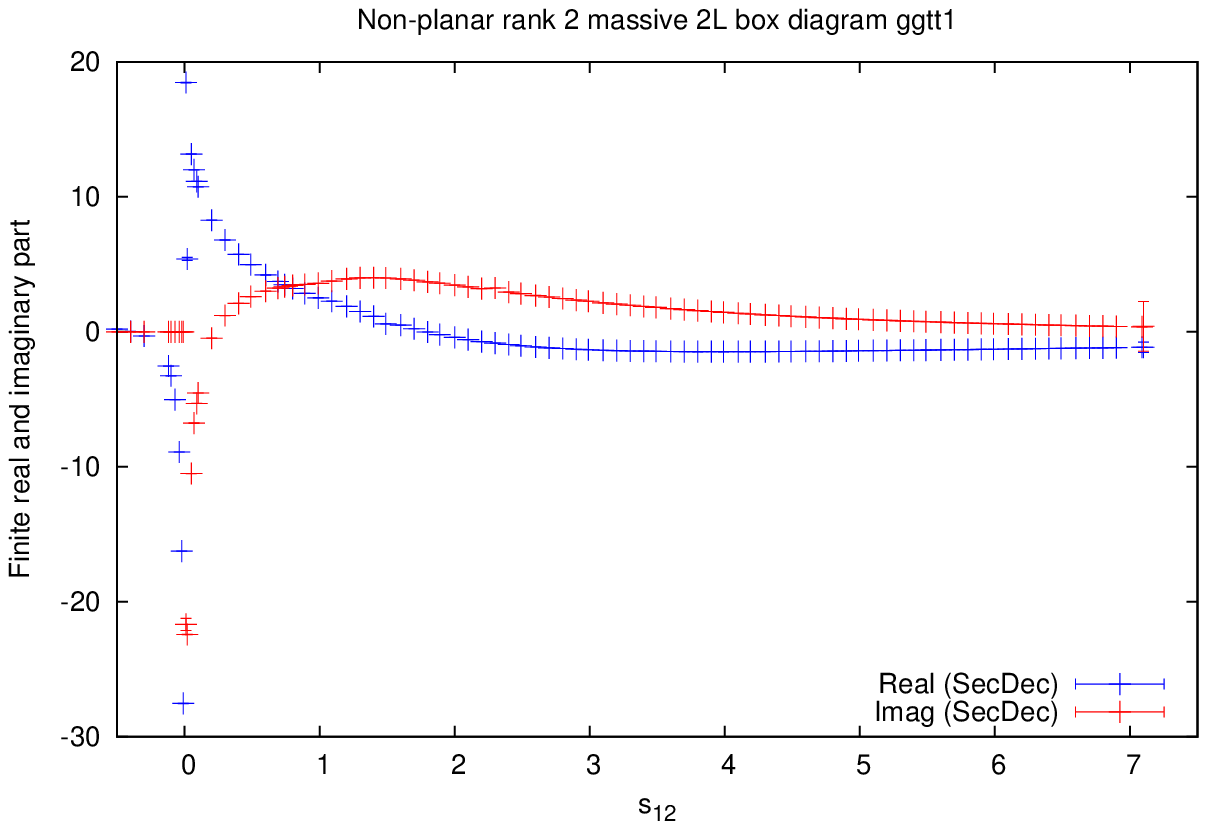} }
\caption{Results for the scalar integral  {\it ggtt1} shown in Fig.~\ref{fig:ggtt2}(a), 
and the corresponding rank two tensor integral {\it ggtt1} with $k_1\cdot k_2$ in the numerator. 
We vary $s_{12}$ and fix $s_{23}=-1.25, m_2=m_1, p_3^2=p_4^2=m_1^2=1$.
}
\label{fig:ggtt1}
 \end{figure}

Numerical results for the diagram {\it ggtt1} (see Fig.~\ref{fig:ggtt2}(a)) are shown in 
Fig.~\ref{fig:ggtt1} for both the scalar integral and an irreducible
rank two tensor integral. 
For the results shown in  Fig.~\ref{fig:ggtt1} we used 
the numerical values $m_1^2=m_2^2=m^2=1, s_{23}=-1.25, s_{13}=2\,m^2-s_{12}-s_{23}$.
We set $m_1^2=m_2^2$ for the results shown in Fig.~\ref{fig:ggtt1} 
because this is the only case occurring in the process $gg\to t\bar{t}$ at two loops
if the $b$-quarks are assumed to be massless. 

The results shown in Fig.~\ref{fig:ggtt1}(b) 
correspond to the rank two  tensor integral with 
a factor of $k_1\cdot k_2$ in the numerator. The numerical integration errors are 
shown as horizontal markers on the vertical lines. The absence of such markers means that 
the numerical errors are smaller than visible in the plot.

The timings for one kinematic point for the scalar integral in Fig.\,\ref{fig:ggtt1}(a) 
range from 11-60 secs for points far from threshold to $1.6\times 10^3$ 
seconds for a point very close to threshold, with an average of about 500 secs 
for points in the vicinity of the threshold.
A relative accuracy 
of $10^{-3}$ has been required for the numerical integration, 
while the absolute accuracy has been set to  $10^{-5}$. 
For the tensor integral, the timings are  better than in the scalar case, 
as the numerator function present in this case smoothes out the singularity structure. 
The timings were obtained on a single machine using Intel i7 processors and 8 cores.

\subsection{Massive tensor two-loop two-point functions}

Here we show that the option to evaluate integrals with a non-trivial numerator 
can also be applied to calculate two-loop two-point functions involving different 
mass scales, without the need 
for a reduction to master integrals. This fact can be used for instance to calculate 
two-loop corrections to mass parameters in a straightforward way.
 
\begin{figure}[htb]
\subfigure[scalar integral]
{\includegraphics[width=0.5\textwidth]{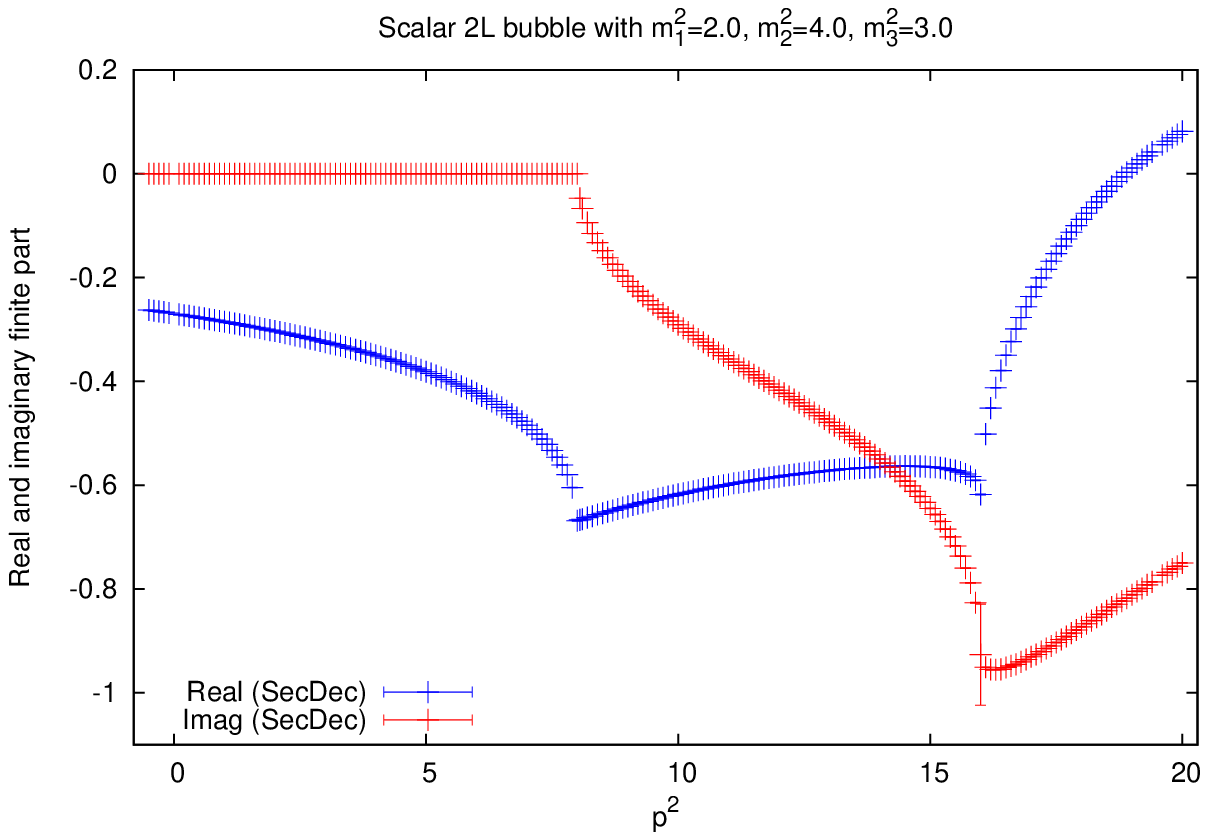}} \hfill
\subfigure[rank 3 tensor integral]
{\includegraphics[width=0.5\textwidth]{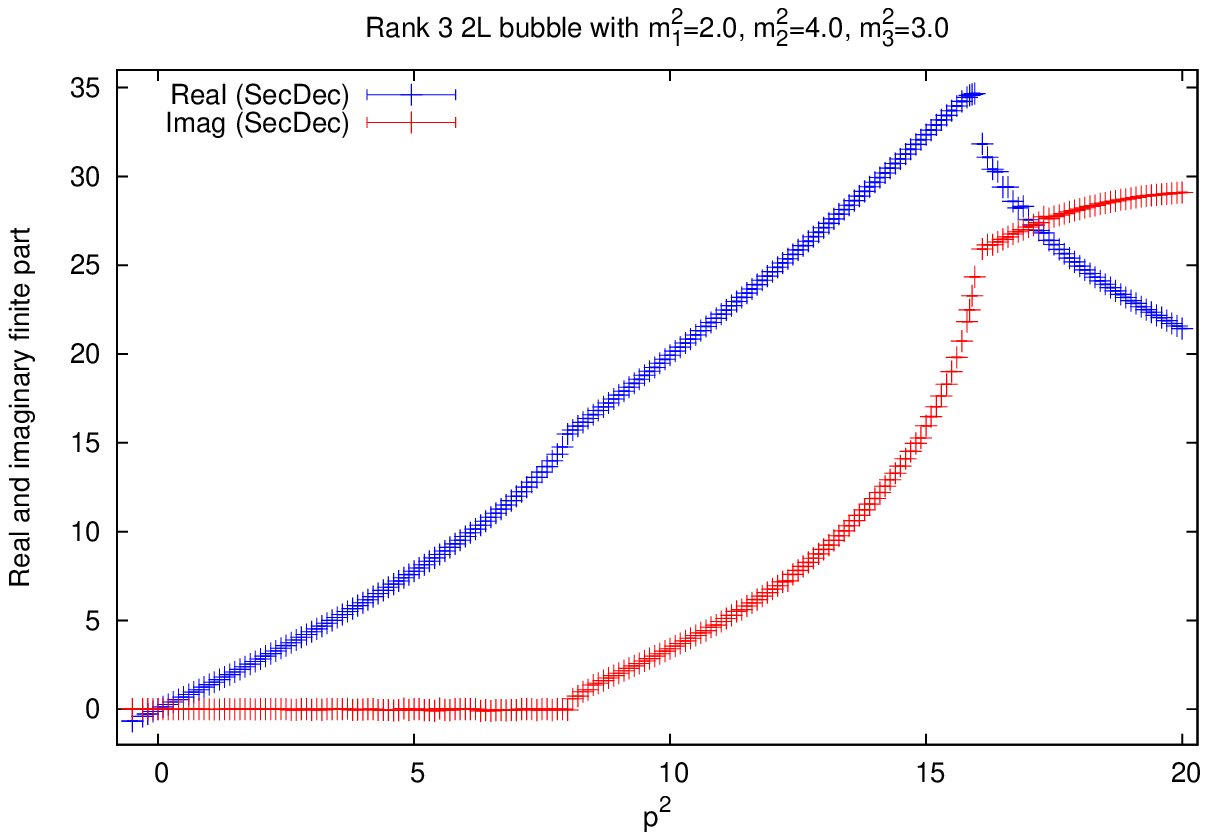} }
\caption{Results for the rank 3 two-loop bubble diagram with three non-vanishing masses.
(a) scalar case, (b) tensor case with numerator $(k_1\cdot k_2)\,(k_1\cdot p_1)$. 
The masses are $m_1^2=2, m_2^2=4, m_3^2=3$.} 
\label{fig:bubble3m} 
\end{figure}

As an example we give results for a scalar integral and a rank three tensor integral, 
where the tensor integral is given by
\begin{align}
\mathcal{G}_{B} &= \left( \frac{1}{\mathrm{i} \pi^{\frac{\mathrm{D}}{2}}}\right)^2 
 \int \frac{\mathrm{d^D}k_1\, \mathrm{d^D}k_2\,\, (k_1\cdot k_2)\,(k_1\cdot p_1)}{D_1\ldots D_5}\;\,; \,
\label{eq:bubbleint}\\
& D_1=k_1^2-m_1^2, D_2=(k_1+p_1)^2-m_1^2, D_3=(k_1-k_2)^2-m_3^2, \nn\\
& D_4=(k_2+p_1)^2-m_2^2, D_5=k_2^2-m_2^2 \;.\nn
\end{align}
The fact that this tensor integral is reducible does not play a role here, because our purpose is
to demonstrate that reduction may become obsolete considering the short integration times for the tensors.
Results  are shown in Fig.~\ref{fig:bubble3m}.

\section{Conclusions}

We have presented  new features of the program \textsc{SecDec}, which can be used to 
calculate multi-loop integrals numerically in an automated way. 
Among the new features are the option to directly calculate tensor integrals,
and to deal with more general 
types of integrands than the ones coming directly from standard Feynman integrals.
These integrands can be defined by the user, while still profiting from 
automated processing for the factorisation of the poles and for the numerical 
integration.

The program is publicly available at {\tt http://projects.hepforge.org/secdec}.

\subsection*{Acknowledgments}
We would like to thank Andreas von Manteuffel for comparisons with analytic results, 
and the organizers of  ACAT2013 for the nice conference.

\section*{References}

\providecommand{\newblock}{}

\end{document}